\documentclass[12pt,aps,prb,preprint]{revtex4}   
\usepackage{amsmath}    % need for subequations
\usepackage{amsfonts}  
\usepackage{amssymb}
\usepackage{graphicx}   % for figures

\begin{document}

\title{Particle in a box with a $\delta$-function potential: strong and weak 
coupling limits}

\author{Yogesh N. Joglekar}
\affiliation{Department of Physics, Indiana University Purdue University 
Indianapolis (IUPUI), Indianapolis, Indiana 46202}
\email{yojoglek@iupui.edu}   %optional
\date{\today}

\begin{abstract}
A particle in a one-dimensional $\delta$-function potential and particle in a 
box are two well-known pedagogical examples; their combination, particle 
in a box with a $\delta$-function potential 
$V_\lambda(x)=\lambda\delta(x-x_0)$, too, has been recently explored. We 
point out that it provides a unique example that is solvable in the weak 
($\lambda\rightarrow 0^{\pm}$) and the strong 
($1/\lambda\rightarrow 0^{\pm}$) coupling limits. In either limit, the 
attractive and repulsive potentials lead to {\it identical spectra}, 
with the possible exception of a single negative-energy state that is 
present when $1/\lambda\rightarrow 0^{-}$. We numerically obtain the spectra 
near the strong-coupling limit and discuss the consequences of the degeneracy 
that arises when $1/\lambda\rightarrow 0^{\pm}$.
\end{abstract}

\maketitle

%-----------------------------------------------------------------------------%

\section{Introduction}
\label{sec:intro}
A particle in a $\delta$-function potential 
$V_\lambda(x)=\lambda\delta(x-x_0)$ and particle in a box of size $a$ that 
runs from $x=0$ to $x=a$ are two pedagogical examples discussed in 
introductory quantum mechanics.~\cite{qm} The first is often used to model 
short-ranged, elastic impurities whereas the second serves as a model for 
semiconductor quantum dots and quantum wells at low temperatures. In the 
first case, when $\lambda<0$, the spectrum develops a single bound state with 
negative energy $E_b=-\lambda^2 m/2\hbar^2$ where $m$ is the mass of the 
particle, and the continuous spectrum at positive energies remains unchanged. 
In the second case, all states are localized within the box and have discrete 
energy eigenvalues given by $E_n=(n\pi)^2E_0$ where $E_0=\hbar^2/2ma^2$ is 
the characteristic energy scale for the box. The problem of particle in a box 
with a $\delta$-function potential has been recently investigated~\cite{jkb} 
using perturbative expansion in the strength of the $\delta$-function 
potential $\lambda$. One salient feature of this problem is that the 
perturbation affects the energies of all eigenfunctions that do not vanish 
at $x_0$; this is in marked contrast to a particle in $\delta$-function 
potential.~\cite{shp}

We point out in this note that the aforementioned problem is solvable in 
both, weak coupling $\lambda\rightarrow 0^{\pm}$ and strong coupling 
$1/\lambda\rightarrow 0^{\pm}$, limits. In either limit {\it the attractive 
and repulsive potentials have identical spectra} except for a single 
bound-state that appears for the attractive potential; identical spectra are 
naturally expected when the $\delta$-function perturbation vanishes, 
$\lambda\rightarrow 0^{\pm}$. The strong-coupling result raises a question 
regarding the completeness of eigenfunctions in these two cases; the 
attractive potential has one more eigenstate - the bound state - 
in the spectrum compared to the repulsive potential. We show that the 
contribution of the bound state to the completeness relation vanishes when 
$1/\lambda\rightarrow 0^{-}$. We numerically obtain the spectra for 
intermediate values of $|\lambda|$ and compare them with perturbative 
corrections to the strong-coupling results.   

%-----------------------------------------------------------------------------%

\section{Particle in a box with a $\delta$-function potential}
\label{sec:one}

Let us consider a particle in a box with $\delta$-function potential inside 
it, $V_\lambda(x)=\lambda\delta(x-x_0)=\lambda\delta(x-pa)$ where 
$0\leq p\leq 1$. The time-dependent Schr\"{o}dinger equation implies that 
the eigenvalues $E_n$ and eigenfunctions $\psi_n(x)$ satisfy
\begin{equation}
\label{eq:sch}
-\frac{\hbar^2}{2m}\frac{d^2}{dx^2}\psi_n(x)+\lambda\delta(x-pa)\psi_n(x)=
E_n\psi_n(x).
\end{equation}
The eigenfunctions $\psi_n(x)$ are continuous and vanish outside the box. At 
positive energies, the (unnormalized) eigenfunctions are given by 
$\psi_n(x)=\sin(k_nx_<)\sin[k_n(a-x_>)]$ where $x_{<(>)}$ is the smaller 
(greater) of ($x, x_0$), and the eigenenergies are $E_n=(k_na)^2 E_0>0$. When 
$E<0$ the corresponding eigenfunction is 
$\psi_\kappa(x)=\sinh(\kappa x_<)\sinh[\kappa(a-x_>)]$ and 
$E=-(\kappa a)^2E_0$. Integrating Eq.(\ref{eq:sch}) over a small interval 
($x_0-\epsilon,x_0+\epsilon$) gives the quantization conditions
\begin{eqnarray}
\label{eq:discont1}
u_n\sin(u_n)+\Lambda\sin(pu_n)\sin[(1-p)u_n]& = & 0,\\
\label{eq:discont2}
v\sinh(v)+\Lambda\sinh(pv)\sinh[(1-p)v]& =& 0,
\end{eqnarray}
where we have defined (dimensionless) $u_n=k_na$, $v=\kappa a$, and the 
dimensionless $\delta$-function strength 
$\Lambda=2ma\lambda/\hbar^2=\lambda/(E_0 a)$. 

\begin{figure}[h]
\begin{center}
\includegraphics{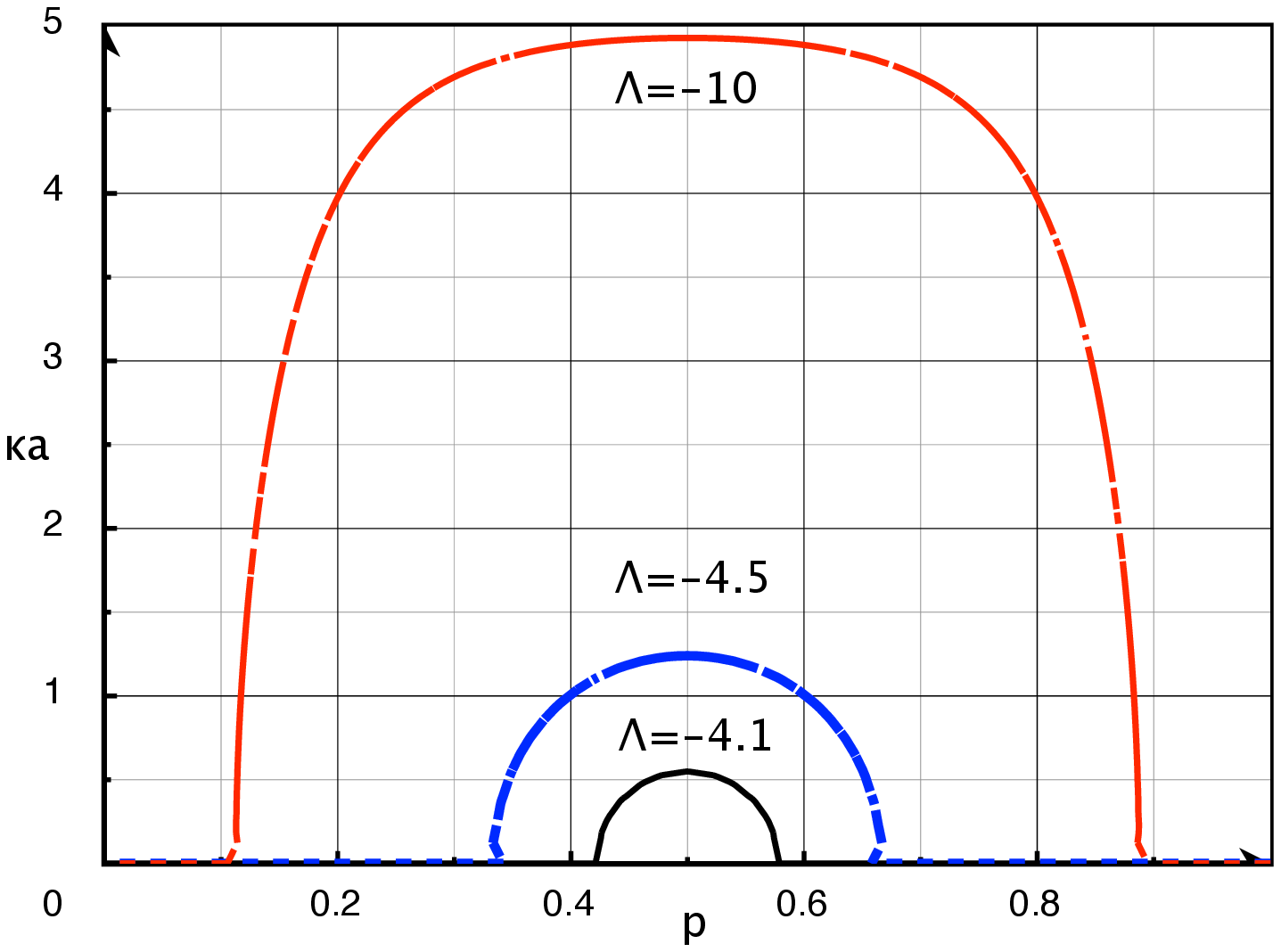}
\caption{\label{fig:bound}The spectrum of the negative-energy state 
$\kappa(p)$ for different strengths of the attractive potential. For 
$|\Lambda|\geq 4$, a single state occurs in the interval of width 
$\Delta p=\sqrt{1-4/|\Lambda|}$ around $p=1/2$. Note that, as discussed 
in the text, $\kappa a(p)\rightarrow|\Lambda|/2$ for large $|\Lambda|$ over 
the interval where $|\Lambda|\gg 1/p(1-p)$.} 
\end{center}
\end{figure}

First we will discuss the negative-energy solution. Eq.(\ref{eq:discont2}) 
has no nonzero solution if $\lambda>0$. For $\lambda<0$, a small-$v$ 
and large-$v$ expansion shows that it has exactly one nonzero solution when 
$|\lambda|>\lambda_c(p)=(E_0a)/p(1-p)$. The critical strength $\lambda_c(p)$ 
required for the negative-energy state increases as the $\delta$-function is 
moved closer to one of the walls. For a ``strong'' attractive potential, 
$|\lambda|\gg\lambda_c(p)$, we recover the result for a free-particle with 
$\delta$-function perturbation, $\kappa a=-\Lambda/2$ and 
$E=-\lambda^2 m/2\hbar^2$. Fig.~\ref{fig:bound} shows the spectra 
$\kappa(p)$ for different strengths of the attractive potential obtained 
by numerically solving Eq.(\ref{eq:discont2}), and verifies the results we 
have derived analytically. 

\begin{figure}
\begin{center}
\includegraphics{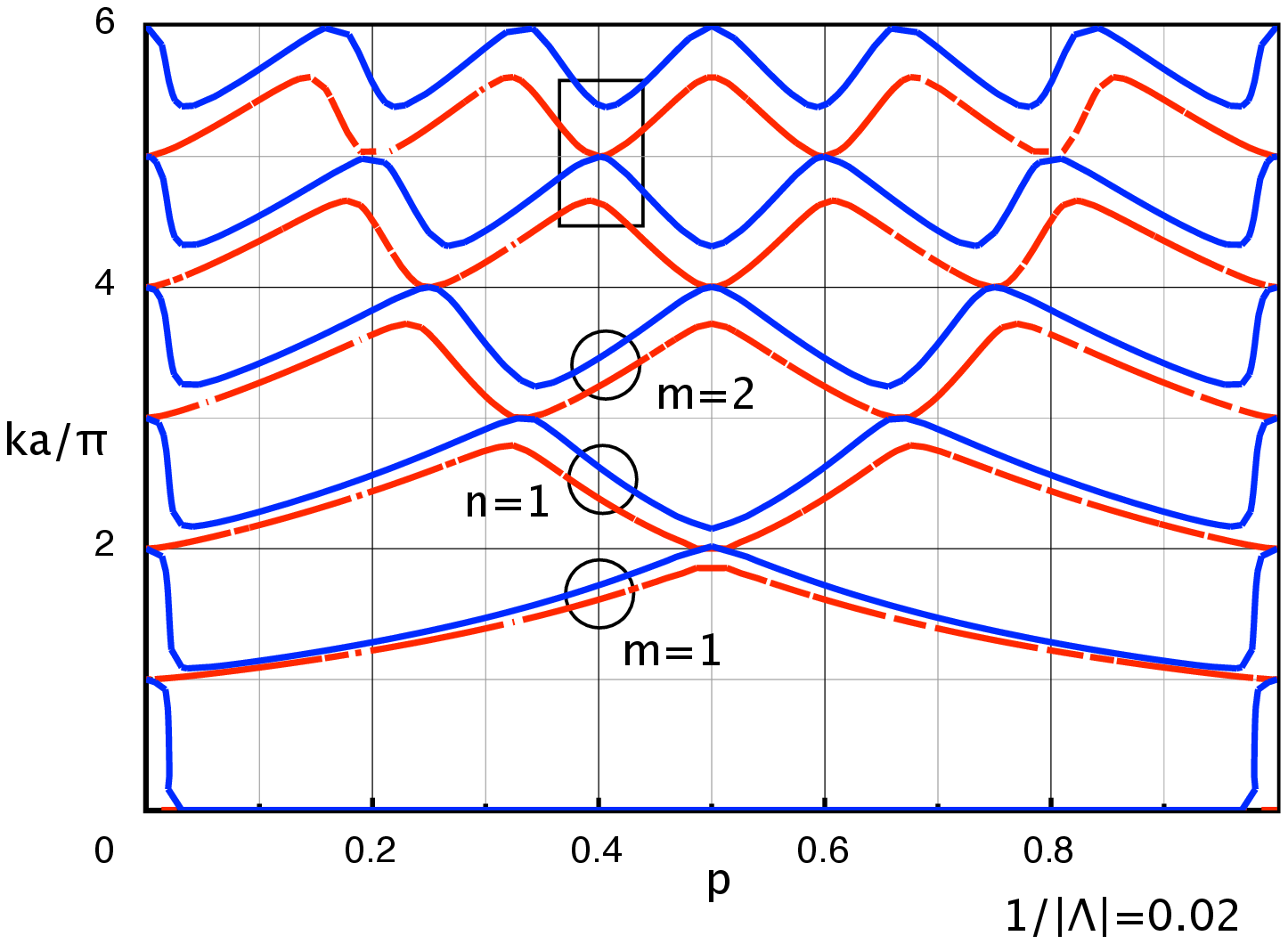}
\caption{\label{fig:spectrum}The spectra $k_\nu(p)$ for attractive 
(blue-solid) and repulsive (red-dotted) potential in the strong coupling 
limit, $1/|\Lambda|=0.02$. Apart from the $j=1$ state pulled down to negative 
energy~\cite{caveat} (bottom-blue-solid), we see that the $(j+1)$-state for 
$\Lambda<0$ (blue-solid) and the $j$-state for $\Lambda>0$ (red-dotted) 
become degenerate as $1/|\Lambda|\rightarrow 0$. The circles show the 
$m=1$, $n=1$, and $m=2$ states from the $p=2/5$ spectrum. The $m=3$ and 
$n=2$ states, shown in the rectangle, become doubly-degenerate in the 
strong-coupling limit $1/|\Lambda|\rightarrow 0$.}
\end{center}
\end{figure}

Next we will focus on the (more interesting) positive energy solutions. When 
$E>0$ the eigenvalues $u_n$ are determined by Eq.(\ref{eq:discont1}). In 
the weak coupling limit $\lambda\rightarrow 0^{\pm}$ we recover the 
well-known result, $k_n=n\pi/a$. In the strong coupling limit 
$1/\lambda\rightarrow 0^{\pm}$, the solutions of Eq.(\ref{eq:discont1}) are 
given by 
\begin{equation}
\label{eq:pboxdelta}
k_\nu(p)=\left\{\frac{n\pi}{ap},\frac{m\pi}{a(1-p)}: m,n=1,2,\ldots\right\}.
\end{equation}
This spectrum is the {\it same irrespective of the sign of the 
potential} and it is symmetric in $p\leftrightarrow(1-p)$. We note 
that the strong-coupling limit corresponds to two infinite wells with 
widths $pa$ and $(1-p)a$ respectively. Figure~\ref{fig:spectrum} shows 
the spectra for both attractive (blue-solid) and repulsive (red-dotted) 
potential when $1/|\Lambda|=0.02$. Recall that the spectrum for particle in 
a box is horizontal lines at $k_\nu(p)=n\pi/a$. In the strong-coupling 
limit, we see that the $(j+1)$-state for attractive potential (blue-solid) 
and the $j$-state for the repulsive potential (red-dotted) approach each 
other.~\cite{caveat} To better understand Eq.(\ref{eq:pboxdelta}), let us 
consider the spectrum for a specific case, say $p=2/5$. 
In Fig.~\ref{fig:spectrum}, the low-lying states $m=1$, $n=1$, $m=2$ are 
marked by the circles. The rectangle in Fig.~\ref{fig:spectrum} 
shows the states with $m=3$ and $n=2$. Since either gives the same value of 
$k_\nu$, the degeneracy in the strong-coupling limit is doubled. In general, 
a double-degeneracy at $k_\nu=N\pi/a$ ($N\geq 2$) arises when $p=\alpha/N$ 
($\alpha=1,\ldots,N-1$). In particular, at the symmetric point $p=1/2$ the 
entire spectrum, given by $k_n=2n\pi/a$, is doubly-degenerate 
and represents the symmetric and antisymmetric states in a double quantum 
well. For a finite $1/\lambda$, Eqs.(\ref{eq:discont1}) and 
(\ref{eq:pboxdelta}) give the following perturbative correction~\cite{caveat3}
\begin{equation}
\label{eq:corr}
k_\nu(p,1/\lambda)=\frac{n\pi}{ap}\left[1-\frac{1}{\Lambda p}\right]
\end{equation}
when $k_\nu(p)=n\pi/ap$ and a corresponding expression with 
$p\leftrightarrow (1-p)$ provided $k_\nu(p)=m\pi/a(1-p)$. Eq.(\ref{eq:corr}) 
shows that near the strong-coupling limit, the repulsive 
potential {\it suppresses} the energy and attractive potential raises 
it. This is in stark contrast with the weak-coupling 
limit~\cite{jkb} where the first order perturbative correction is given by 
\begin{equation}
k_n(p,\lambda)=\frac{n\pi}{a}\left[1+\Lambda\frac{\sin^2(n\pi x_0/a)}
{(n\pi)^2}\right].
\end{equation} 
This unusual behavior arises because, in contrast to all other potentials, 
the $\delta$-function spectrum is well-defined in the strong-coupling limit, 
and is the same irrespective of the sign of the potential. 

We conclude the note with a comment on the completeness relation. The 
completeness of eigenfunctions in the attractive and repulsive cases implies 
that 
\begin{eqnarray}
\label{eq:1positive}
\sum_{\nu}\phi_{k_\nu}(x)\phi^{*}_{k_\nu}(x')=\delta(x-x') &&(\lambda>0),\\
\label{eq:1negative}
\phi_{\kappa}(x)\phi_{\kappa}(x')+
\sum_{\nu}\phi_{k_\nu}(x)\phi^{*}_{k_\nu}(x')=\delta(x-x') &&(\lambda<0).
\end{eqnarray}
where $0\le x,x'\le a$, $\phi_{k_\nu}$ are the {\it normalized} 
positive-energy eigenfunctions, and the {\it normalized} negative-energy 
eigenfunction is given by $\phi_\kappa(x)=A\psi_\kappa(x)$ with 
\begin{equation}
\label{eq:norm}
A^{-2} =\frac{1}{4\kappa}\left\{\sinh^{2}[(1-p)\kappa a]
\left[\sinh(2p\kappa a)-
(2p\kappa a)\right]+\left[p\rightarrow (1-p)\right]\right\}.
\end{equation}
In the strong-coupling limit, $\kappa a\gg 1$ and Eq.(\ref{eq:norm}) implies 
that $A\sim 4\sqrt{\kappa}\exp(-\kappa a)$. Therefore, the contribution to 
Eq.(\ref{eq:1negative}) from the negative-energy state vanishes in the 
strong-coupling limit, as the two spectra at positive energies 
converge.~\cite{caveat2}  

%-----------------------------------------------------------------------------%

\begin{acknowledgments}
It is a pleasure to thank Ricardo Decca and Durgu Rao for helpful discussions. 
\end{acknowledgments}

%-----------------------------------------------------------------------------%

%-----------------------------------------------------------------------------%

\newpage
\section*{Figures}

%-----------------------------------------------------------------------------%


\begin{thebibliography}{99}
\bibitem{caveat}In the strong-coupling limit, the ground state ($j=1$) for 
attractive potential is at negative energy, and we do not consider it in 
the spectrum presented in Fig.~\ref{fig:spectrum}.
\bibitem{caveat2} This contribution also vanishes in the weak-coupling limit, 
$\kappa a\ll 1$ and is consistent with the results in Ref.~\onlinecite{shp}.
\end{thebibliography}
\end{document}